\documentclass[preprintnumbers,aps,prl,twocolumn,showpacs,showkeys]{revtex4} 
\usepackage{amsmath,amsfonts,amssymb,amscd,amsxtra,amsthm}
\usepackage{graphicx}
 \usepackage{braket}
\usepackage{epstopdf}
\usepackage{bm}
\begin{document}
\preprint{NuHaTh-2014-01}
\title{Jet-quenching parameter $\hat{q}$ in the nonperturbative region}
%--------------------------------------------------
\author{Seung-il Nam}
\email[E-mail: ]{sinam@pknu.ac.kr}
\affiliation{ Department of Physics, Pukyong National University (PKNU), Busan 608-737, Republic of Korea}
%--------------------------------------------------
\date{\today}
\begin{abstract}
We investigate the jet-quenching parameter $\hat{q}$ for the quark-gluon plasma (QGP) for $N_c=3$, defined nonperturbatively with the Wilson loop in the light-cone (LC) coordinate, at finite temperature $(T)$. Considering the effective static (heavy) quark-antiquark potential $V_{\bar{Q}Q}=\sigma L+C-A/L$, where $L$ indicates the LC transverse separation between the quarks, we obtain $\hat{q}\approx8V_{\bar{Q}Q}/L^2$. The $T$ dependences for $L$ and other relevant parameters are extracted from the $T$-dependent instanton (trivial-holonomy caloron) length parameters and the lattice QCD data. By choosing $L\approx{a}\approx\bar{\rho}_T$, in which $a$ and $\bar{\rho}_T$ denote the lattice spacing and the $T$-dependent average (anti)instanton size, respectively, we acquire numerically that $\hat{q}=(5\sim25)\,\mathrm{GeV/fm}$ for $T=(0\sim0.6)$ GeV, and these values are well consistent with other estimations from AdS/CFT and experimental analyses. It turns out that $\hat{q}$ is produced almost by the Coulomb and constant potentials of $V_{\bar{Q}Q}$. We also observe that the ratio $T^3/\hat{q}$ turns into being saturated to $\sim4.45\times10^{-2}$ for $T\gtrsim0.4$ GeV, indicating the strongly-coupled QGP, and $\hat{q}$ behaves proportionally to $T^3$ for high $T$. 
\end{abstract}
\maketitle
%--------------------------------------------------
%\section{Introduction}
%--------------------------------------------------
The heavy-ion collision experiments, which have been carried out in the Relativistic Heavy-Ion Collider (RHIC) at the Brookhaven National Laboratory (BNL) and the Large Hadron Collider (LHC) of the European Organization for Nuclear Research (CERN) for instance, are a unique place to explore the various nontrivial phenomena of Quantum ChromoDynamics (QCD). It has been reported that a highly hot and dense matter is produced during the very short time after the collision: The quark-gluon plasma (QGP). From the comparison between theory of the relativistic hydrodynamic calculation and experiments~\cite{Song:2010mg} for the Fourier coefficient of the elliptic flow, the QGP is now believed as an almost perfect fluid, being indicated by the small shear viscosity-to-entropy density ratio $\eta/s\ge1/(4\pi)$, known as the Kovtun-Son-Starinets (KSS) bound~\cite{Kovtun:2004de}. In Ref.~\cite{Abelev:2009ac}, a significant charge separation was observed in the presence of the strong magnetic field, and this was interpreted theoretically in terms of the Chiral Magnetic Effect (CME)~\cite{Fukushima:2008xe}. In addition, it is worth mentioning that there appears a suppression of the hadrons in the high-$p_T$, i.e. medium modification of the parton fragmentation, and this phenomenon can be understood by the so-called {\it jet quenching}, which amounts the radiative energy loss of the multiple scattering of the partons~\cite{Liu:2006he,Majumder:2007zh,Benzke:2012sz,Buchel:2006bv}. Hence, the strength of the jet quenching in the medium can be quantified by defining the Jet-Quenching Parameter (JQP) $\hat{q}$. This physical quantity is also employed as a physical parameter in the hydrodynamic simulations for the heavy-ion collisions.  We take notice of that $\hat{q}$ has been investigated theoretically and phenomenologically from various points of view so far. A perturbative definition for $\hat{q}$ was suggested in Ref.~\cite{Majumder:2007zh} in terms of the light-cone correlation of the gluon field strength tensors. In that work, the ratio of $T^3/\hat{q}$ was taken into consideration as a more broadly-valid measure than $\eta/s$, and gives a criteria that $\eta/s\,(\approx,\gg)\,1.25\,T^3/\hat{q}$ for the (weakly, strongly) coupled medium.  In Ref.~\cite{Majumder:2007ae}, the three-dimensional fluid dynamics was performed for the most central Au$+$Au collision at the next-to-leading (NLO) twist, resulting in $\hat{q}=(1\sim2)\,\mathrm{GeV}^2$/fm with the initial time $\sim1$ fm and $T\approx400$ MeV. A nonperturbative definition for $\hat{q}$ was developed by employing the vacuum expectation value (VEV) of the closed contour Wilson loop in the light-cone frame in terms of the dipole approximation~\cite{Liu:2006he}. With the nonperturbative definition for $\hat{q}$, several works in the context of the gauge/string duality, also known as the AdS/CFT, computed $\hat{q}$: The hot $\mathcal{N}=4$ supersymmetric QCD~\cite{Liu:2006he}, medium with chemical potential~\cite{Lin:2006au}, and  bulk geometry with the  hyperscaling violation~\cite{Sadeghi:2013dga}. All of these AdS/CFT approaches showed that $\hat{q}=(3\sim20)\,\mathrm{GeV}^2/$fm for $T=(250\sim500)$ MeV, showing the tendency that $\hat{q}$ is proportional to $T^3$. In the present work, we investigate $\hat{q}$ for the quark-gluon plasma (QGP) for $N_c=3$, defined nonperturbatively with the Wilson loop in the light-cone (LC) coordinate, at finite temperature $(T)$. To this end, we will utilize the effective static (heavy) quark-antiquark  potential with help of the lattice QCD (LQCD) information and $T$-dependent instanton model.  
%--------------------------------------------------
%\section{Theoretical Framework}
%--------------------------------------------------

%FIGURE>>>
\begin{figure}[t]
\includegraphics[width=7.5cm]{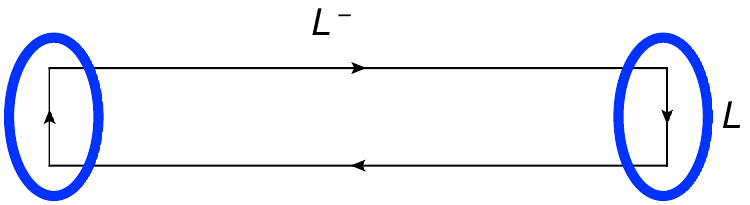}
\caption{Light-cone Wilson loop with the light-cone (LC) distance $L^-$ and LC transverse $\bar{Q}Q$ separation $L$~\cite{Liu:2006he}. The ellipse indicates the spatial distribution of the instanton with its size $\bar{\rho}\approx1/3$ fm.}       
\label{FIG0}
\end{figure}
%FIGURE<<<

\begin{figure}[t]
\begin{tabular}{cc}
\includegraphics[width=4.2cm]{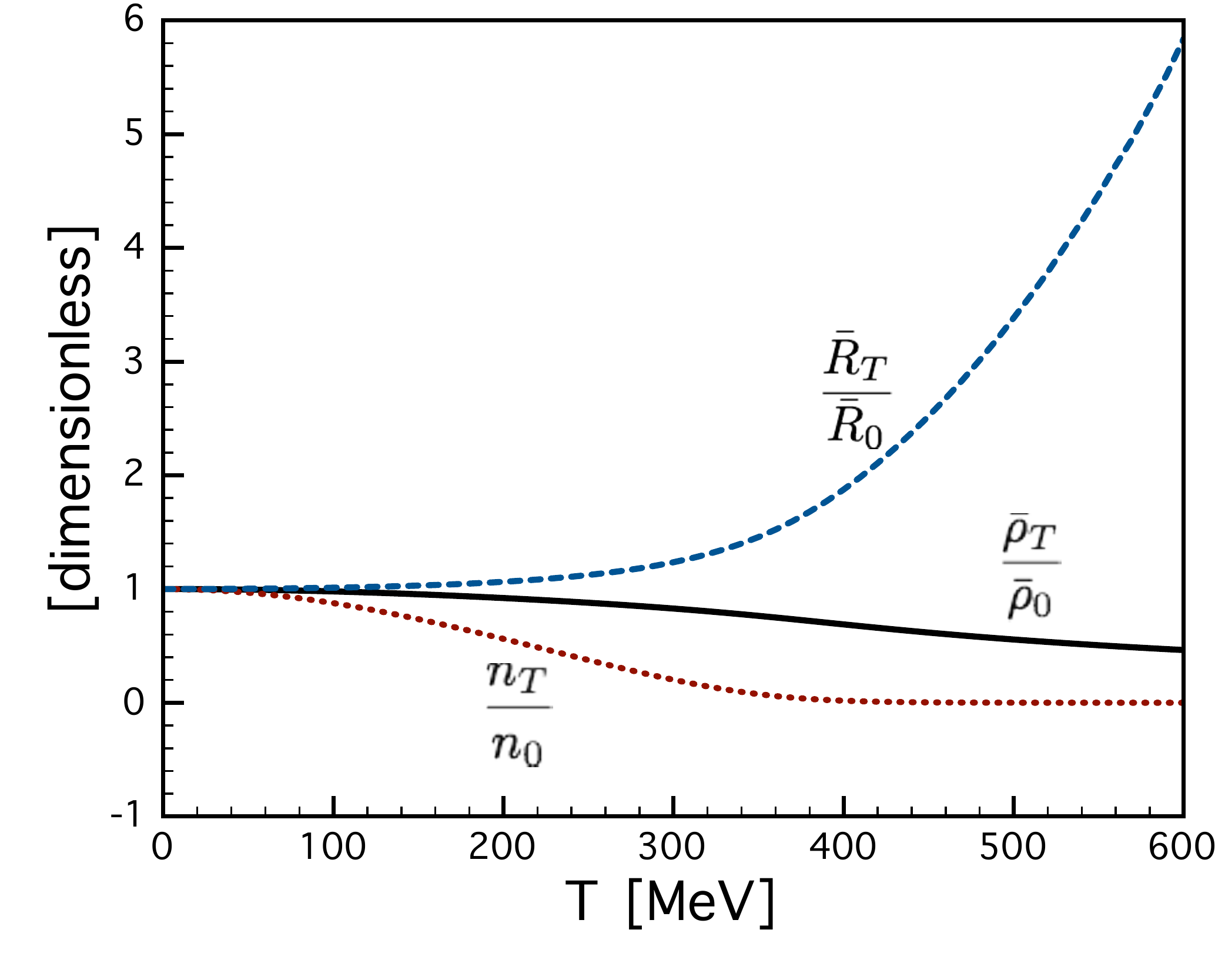}
\includegraphics[width=4.2cm]{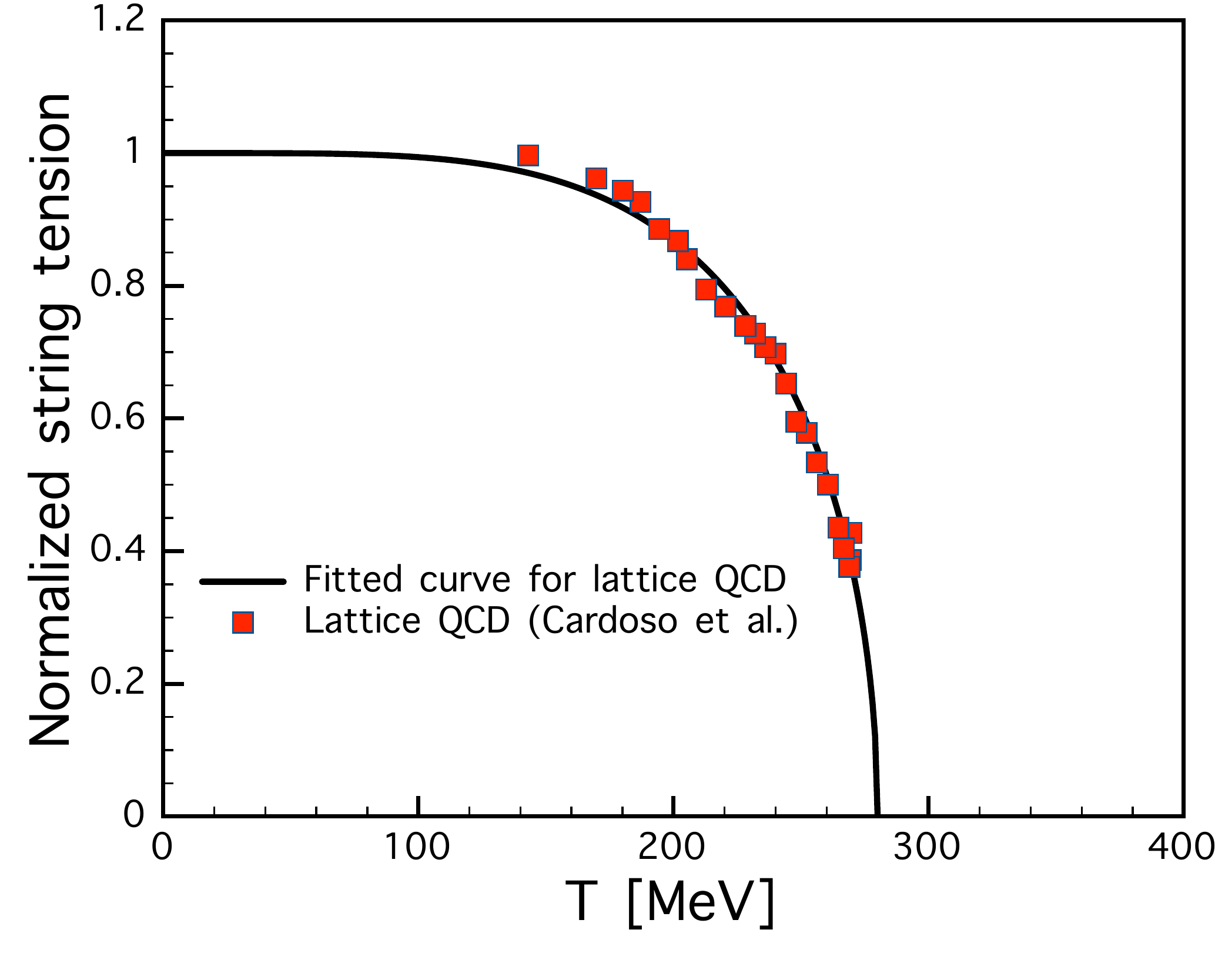}
\end{tabular}
\caption{(Color online) Left: $\bar{\rho}_T/\bar{\rho}_0$ (solid), $\mathcal{G}_T/\mathcal{G}_0$ (dot), and $\bar{R}_T/\bar{R_0}$ as functions of $T$. Right: Normalized string tension from the LQCD simulation~\cite{Cardoso:2011hh} (square), its parameterization (solid).}       
\label{FIG12}
\end{figure}

Firstly, we start by introducing the present theoretical framework for computing $\hat{q}$ in brief. The Wilson loop in a closed contour $\mathcal{C}$ can be described with the number of color $N_c$ as follows:
%EQUATION>>>
\begin{equation}
\label{eq:WILDEF}
W(\mathcal{C})=\frac{1}{N_c}\Big<\mathrm{Tr}\left[\mathcal{P}\exp\left[i\oint_\mathcal{C}g_sA_\mu (x)\,dx^\mu\right] \right]\Big>,
\end{equation}
%EQUAITON<<<
where $\mathcal{P}$, $g_s$ and $A_\mu$ indicate the path-ordering operator, the gauge coupling and field for the SU($N_c$) gauge group. In Ref.~\cite{Liu:2006he}, $\hat{q}$ was defined nonperturbatively in the static medium, in terms of the VEV of the LC Wilson loop in the {\it adjoint} $(A)$ representation, which is closed by the LC contour, i.e. $\mathcal{C}=\mathrm{LC}$, as shown in Fig.~\ref{FIG0}: 
%EQUATION>>>
\begin{equation}
\label{eq:WILSONLIU}
\langle W_A(\mathrm{LC})\rangle
=\exp\left[-\frac{1}{4\sqrt{2}}\hat{q}L^-L^2 \right]+\mathcal{O}\left(\frac{1}{N^2_c} \right).
\end{equation}
%EQUAITON<<<
Here, $L$ and $L^-$ represent the transverse  separation between the (heavy) quarks and the LC distance, respectively, satisfying the hierarchy $L\ll L^-$. Note that $L$ also corresponds to the size of the color dipole. They are related to transverse $(r)$ and longitudinal $(\Delta z)$ distances in the following way~\cite{Liu:2006he}:
%EQUATION>>>
\begin{equation}
\label{eq:LENGTH}
L= \sqrt{\Delta x^2+\Delta y^2}\equiv r,\,\,\,L^-=\sqrt{2}\Delta z.
\end{equation}
%EQUAITON<<<
Being together with this setup, the closed area for the LC Wilson loop can be taken into account as one of the $\frac{\pi}{4}$-tilted Wilson loops in Euclidean space~\cite{Pirner:2004qd}, resulting in 
%EQUATION>>>
\begin{equation}
\label{eq:SSSS}
s_\mathrm{LC}=\frac{1}{\sqrt{2}}LL^-.
\end{equation}
%EQUAITON<<<
Secondly, the VEV of an effective LC wilson loop can be constructed from the LQCD and operator product expansion (OPE) information. It is well-known that the static (heavy) quark-antiquark potential can be approximated as the sum of Coulomb, linear, and constant potentials:
%EQUATION>>>
\begin{equation}
\label{eq:EFFPOT}
V_{Q\bar{Q}}(r)=-\frac{A}{r}+\sigma r+C.
\end{equation}
%EQUAITON<<<
The values for $\sigma$, $A$, and $C$ were estimated using the quenched LQCD as follows~\cite{Takahashi:2000te}:
%EQUATION>>>
\begin{equation}
\label{eq:SAC}
\sigma a^2=0.1629,\,\,\,\,A=0.2793,\,\,\,\,Ca=0.6203.
\end{equation}
%EQUAITON<<<
Here, $a$ and $\sigma$ stand for the lattice spacing and string tension. In Ref.~\cite{Shifman:1980ui}, it was also studied that the Wilson loop can be interpreted with the gluon condensate, which is an important order parameters for the nonperturbative QCD, in terms of the OPE:
%EQUATION>>>
\begin{equation}
\label{eq:WILSONSHIF}
\langle W(\mathcal{C})\rangle=\frac{2}{zs_\mathcal{C}}J_1(zs_\mathcal{C})=1-\frac{z^2s^2_\mathcal{C}}{8}+\cdots,
\end{equation}
%EQUAITON<<<
where  $z^2\equiv \frac{2\pi^2}{3N_c}\mathcal{G}$, in which $\mathcal{G}\equiv\langle \frac{\alpha_s}{\pi}G^2\rangle$ stands for the  gluon condensate. In the vacuum, we have $\mathcal{G}_0\approx(0.33\pm0.04)^4\,\mathrm{GeV}^4$~\cite{Furnstahl:1992pi}. Although  the expression for the Wilson loop in Eq.~(\ref{eq:WILSONSHIF}) does not satisfy the positivity of VEV in the large $z^2s^2_\mathcal{C}$ region, we speculate that the effective Wilson loop follows the asymptotic behavior of Eq.~(\ref{eq:WILSONSHIF}) for $s_\mathcal{C}\to0$. Taking into account the above discussions, from a phenomenological point of view, we suggest an effective Wilson loop as follows in the {\it fundamental} ($F$) representation up to $\mathcal{O}(s^2_\mathcal{C})$ in Euclidean space:
%EQUATION>>>
\begin{equation}
\label{eq:WILSONEFF}
\langle W_F(\mathcal{C})\rangle
=\exp\left[-C \tau-\left(\sigma-\frac{A}{  r^2} \right) s_\mathcal{C}-\frac{z^2s^2_\mathcal{C}}{8}\right],
\end{equation}
%EQUAITON<<<
where $ \tau  r=s_\mathcal{C}$. Eq.~(\ref{eq:WILSONEFF}) satisfies that $\langle W_F(\mathcal{C})\rangle$ becomes unity $ \tau\to0$, and $-\lim_{\tau\to\infty}\frac{1}{\tau}\ln[\langle W_F(\mathcal{C})\rangle]=V_{\bar{Q}Q}$ for the small $s_\mathcal{C}$. Using Eqs.~(\ref{eq:LENGTH}) and (\ref{eq:SSSS}), and considering that $\ln\langle W_A(\mathcal{C})\rangle=2\ln\langle W_F(\mathcal{C})\rangle$~\cite{Liu:2006he}, we can rewrite Eq.~(\ref{eq:WILSONEFF}) in the LC coordinate as follows:
%EQUATION>>>
\begin{eqnarray}
\label{eq:WILSONEFFLC}
\langle W_A(\mathrm{LC})\rangle&=&
\exp\Bigg[-\sqrt{2}CL^--\sqrt{2}\left(\sigma -\frac{A}{L^2} \right)L^-L
\cr
&-&\frac{z^2}{8}( L^-L)^2\Bigg].
\end{eqnarray}
%EQUAITON<<<
Then, by equating Eqs.~(\ref{eq:WILSONLIU}) and (\ref{eq:WILSONEFFLC}) up to leading large-$N_c$ contributions and $\mathcal{O}(s^2_\mathrm{LC})$, one is led to 
%EQUATION>>>
\begin{equation}
\label{eq:JQP}
\hat{q}=\sum^3_{n=0}\hat{q}_n\left(\frac{1}{L^n} \right)=\frac{1}{\sqrt{2}}z^2L^-+\frac{8\sigma}{L}+\frac{8C}{L^2}- \frac{8A}{L^3}.
\end{equation}
%EQUAITON<<<
As will be shown later, since the contribution from the first term $\hat{q}_0\propto L^-$ is relatively negligible, by comparing Eq.~(\ref{eq:EFFPOT}) with Eq.~(\ref{eq:JQP}), we obtain a simple and useful relation between $\hat{q}$ and $V_{\bar{Q}Q}$:
%EQUATION>>>
\begin{equation}
\label{eq:JQPSIM}
\hat{q}\approx\frac{8V_{\bar{Q}Q}}{L^2}.
\end{equation}
%EQUAITON<<<
Since we are interested in computing $\hat{q}$ at finite $T$, it is necessary to modify all the relevant variables, $z$, $L^-$, $L$, $\sigma$, and so on  in Eq.~(\ref{eq:JQP}) as functions of $T$. For this purpose, we want to employ the LQCD data and the modified liquid-instanton model (mLIM). In this model, which is properly defined in Euclidean space, all the relevant instanton parameters, such as the average inter-(anti)instanton distance $\bar{R}$ and average (anti)instanton size ($\bar{\rho}$), are given as functions of $T$, using the trivial caloron solution, i.e. Harrington-Shepard caloron, as the compactified instanton solution through the Euclidean temporal direction~\cite{Harrington:1976dj,Diakonov:1988my,Nam:2009nn}. An instanton distribution function for arbitrary $N_c$ can be written with a Gaussian suppression factor as a function of $T$ and an arbitrary instanton size $\rho$ for the pure-glue QCD, i.e. Yang-Mills equation in Euclidean space~\cite{Diakonov:1988my}:
%EQUATION>>>
\begin{eqnarray}
\label{eq:INSDIS}
d(\rho,T)&=&\mathcal{C}\,\rho^{b-5}
\exp\left[-\mathcal{F}(T)\rho^2 \right],
\cr
\mathcal{F}(T)&=&\frac{1}{2}A_{N_c}T^2+\left[\frac{1}{4}A^2_{N_c}T^4
+\frac{\nu\bar{\beta}\gamma}{\bar{R}^4} \right]^{\frac{1}{2}}.
\end{eqnarray}
%EQUAITON<<<
Note that the $CP$-invariant vacuum was taken into account in Eq.~(\ref{eq:INSDIS}), and we assumed the same analytical form of the distribution function for the (anti)instanton. The instanton packing fraction $1/\bar{R}^4$ and $\bar{\rho}$ are functions of $T$ implicitly here. The notations read for $N_c=3$ as follows:
%EQUATION>>>
\begin{equation}
\label{eq:PARA}
A_{N_c}=\frac{3}{2}\pi^2,\,\,\gamma=\frac{81}{32}\pi^2,\,\,b=11,\,\,\nu=\frac{7}{2}.
\end{equation}
%EQUAITON<<<
Using the instanton distribution function in Eq.~(\ref{eq:INSDIS}), we can compute the average value of the instanton size $\bar{\rho}^2$ straightforwardly as a function of $T$ as follows~\cite{Schafer:1996wv}:
%EQUATION>>>
\begin{equation}
\label{eq:rho}
\bar{\rho}^2_T\equiv\bar{\rho}^2(T)=\frac{\left[A^2_{N_c}T^4
+4\nu\bar{\beta}\gamma n \right]^{\frac{1}{2}}
-A_{N_c}T^2}{2\bar{\beta}\gamma n}.
\end{equation}
%EQUATION<<<
It is clear that Eq.~(\ref{eq:rho}) satisfies the following asymptotic behavior~\cite{Schafer:1996wv}: 
%EQUATION>>>
\begin{equation}
\label{eq:RHOHT}
\lim_{T\to\infty}\bar{\rho}^2_T=\frac{\nu}{A_{N_c}T^2},
\end{equation}
%EQUAITON<<<
which shows a correct scale-temperature behavior at high $T$, i.e., $1/\bar{\rho}\propto T$, since $1/\bar{\rho}$ is characterized as the renormalization scale of mLIM: $\Lambda_\mathrm{mLIM}\approx600$ MeV in the vacuum~\cite{Diakonov:1988my}. We can also compute the instanton packing fraction $1/\bar{R}^4$ or the instanton number density $n$ by solving the following equation:
%EQUATION>>>
\begin{equation}
\label{eq:NOVV}
n^{\frac{1}{\nu}}\mathcal{F}(T)=\left[\mathcal{C}\,\Gamma(\nu) \right]^\frac{1}{\nu},
\end{equation}
%EQUATION<<<s
where $\Gamma(\nu)$ stands for the Gamma function with the argument $\nu$. Note that $1/{\bar{R}^4}=n$ is a function of $T$ implicitly: $\bar{R}_T\equiv \bar{R}(T)$ and $n_T\equiv n(T)$. By equating Eqs.~(\ref{eq:INSDIS}) and (\ref{eq:NOVV}), we have $\bar{\beta}=\nu/(\gamma\bar{\rho}^4n)$. For simplicity, we determine the value of $\bar{\beta}$ at $T=0$: $\bar{\beta}=\nu/(\gamma\bar{\rho}^4_0n_0)$. Similarly, the coefficient $\mathcal{C}$ becomes $n_0(\nu\bar{\beta}\gamma n_0)^{\nu/2}/\Gamma(\nu)$. It is worth mentioning that $n$ corresponds to the gluon condensate in a convenient form: $\frac{1}{\bar{R}^4}= n= \frac{1}{8}\mathcal{G}$. Considering all the ingredients above, we can obtain the $T$ dependences of the relevant parameters, i.e. $\bar{\rho}_T$, $n_T$, and $\bar{R}_T$,  numerically, as shown in the left panel of Fig.~\ref{FIG12}. At $T=0$, we have $\bar{\rho}_0=0.345$ fm, $n_0=1.625\times 10^{-3}\,\mathrm{GeV}^4$, and $\bar{R}_0\approx1.020$ fm, giving the gluon condensate $\mathcal{G}_0=0.013\,\mathrm{GeV}^4$, which is well consistent with the phenomenological value~\cite{Furnstahl:1992pi}. Note that here is a {\it hierarchy} in the length parameters $\bar{\rho}_T\ll\bar{R}_T$, and  their discrepancy becomes more obvious as $T$ increases as shown in the figure. The curve for $\bar{\rho}_T$ indicates that the average (anti)instanton size smoothly decreases with respect to $T$, and inversely for $\bar{R}_T$. This tendency shows that the instanton ensemble becomes diluted and the nonperturbative effects from the quark-instanton interactions are diminished. Taking into account that the instanton size corresponds to the renormalization scale of the present model, i.e. UV cutoff mass, $\bar{\rho}\approx1/\Lambda_\mathrm{mLIM}$, the $T$-dependent cutoff mass is a clearly distinctive feature in comparison to other low-energy effective models, such as the NJL model. We also show $n_T/n_0=\mathcal{G}_T/\mathcal{G}_0$ there. As observed, the gluon condensate, $\mathcal{G}_T=8n_T$ decreases with respect to $T$, indicating the partial chiral restoration, being similar to that of $\bar{\rho}_T$. We also note that the gluon condensate small but remains finite beyond $T\approx300$ MeV. Considering that the nonperturbative effects are generated by the instanton contributions and their hierarchies in the length parameters, the LC lengths, $L$ and $L^-$, are assumed to be in the relations that $\bar{\rho}_T\approx L$ and $\bar{R}_T\approx L^-/\sqrt{2}$, using Eq.~(\ref{eq:LENGTH}), as shown in Fig.~\ref{FIG0}. As for the $T$ dependence of the string tension $\sigma_T$, we used it from the LQCD data~\cite{Cardoso:2011hh} as shown by the square in the right panel of Fig~\ref{FIG12}, and parameterize it with  $(\sigma/\sigma_0)^2=1-[(T-T_0)/T_C]^4$ for $(T_0,T_c)=(10,280)\,\mathrm{MeV}$, given by the solid line. Since $\bar{\rho}_T$ relates to the cutoff scale $\sim\Lambda_\mathrm{mLIM}$, it is natural to choose the lattice spacing as $a\approx\bar{\rho}_T$. Due to this choice, the value for $\sigma_0$ is chosen to be $0.27\,\mathrm{GeV/fm}$ with Eq.~(\ref{eq:SAC}) and $a\approx\bar{\rho}_0=0.345$ fm. Note that this value is smaller than the phenomenological one $\sigma_0=0.89\,\mathrm{GeV/fm}$ for $a=0.19$ fm~\cite{Takahashi:2000te}. It is also interesting to see the asymptotic behavior of $\hat{q}$ as $T\to\infty$. Using Eqs.~(\ref{eq:JQP}) and (\ref{eq:RHOHT}), and ignoring $\hat{q}_0$, then we have a simple expression as follows:
%EQUATION>>>
\begin{equation}
\label{eq:PARAJQP}
\hat{q}_{\infty}\approx8V_{\bar{Q}Q}\pi^3
\left(\frac{11N_c-6}{33N_c-36} \right)^{\frac{3}{2}}T^3,
\end{equation}
%EQUAITON<<<
which gives $\hat{q}_{\infty}\approx177.743\times T^3\,[\mathrm{GeV/fm}]$ for $N_c=3$ with Eq.~(\ref{eq:SAC}).

%FIGURE>>>
%FIGURE<<<

\begin{figure}[t]
\begin{tabular}{cc}
\includegraphics[width=4.2cm]{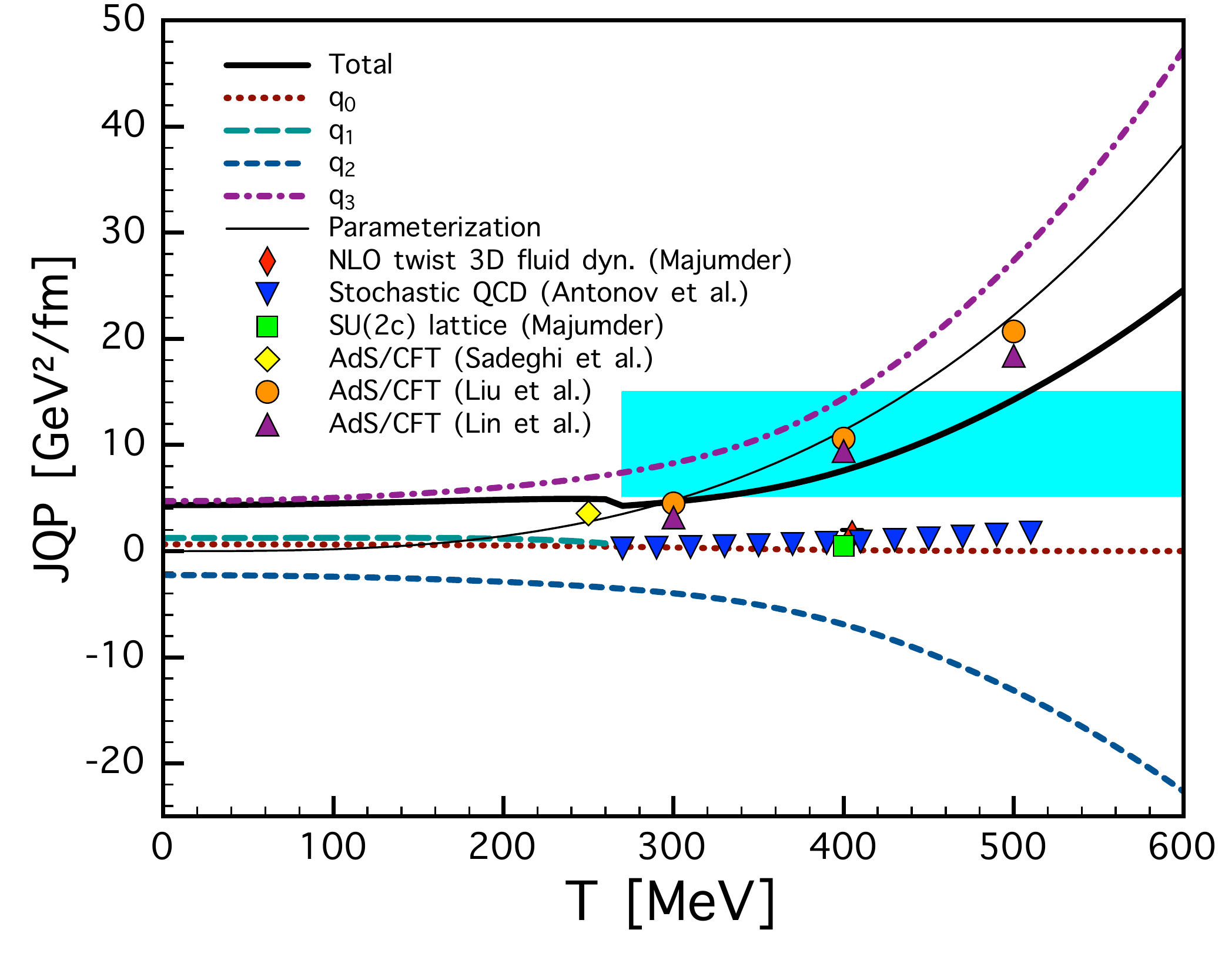}
\includegraphics[width=4.2cm]{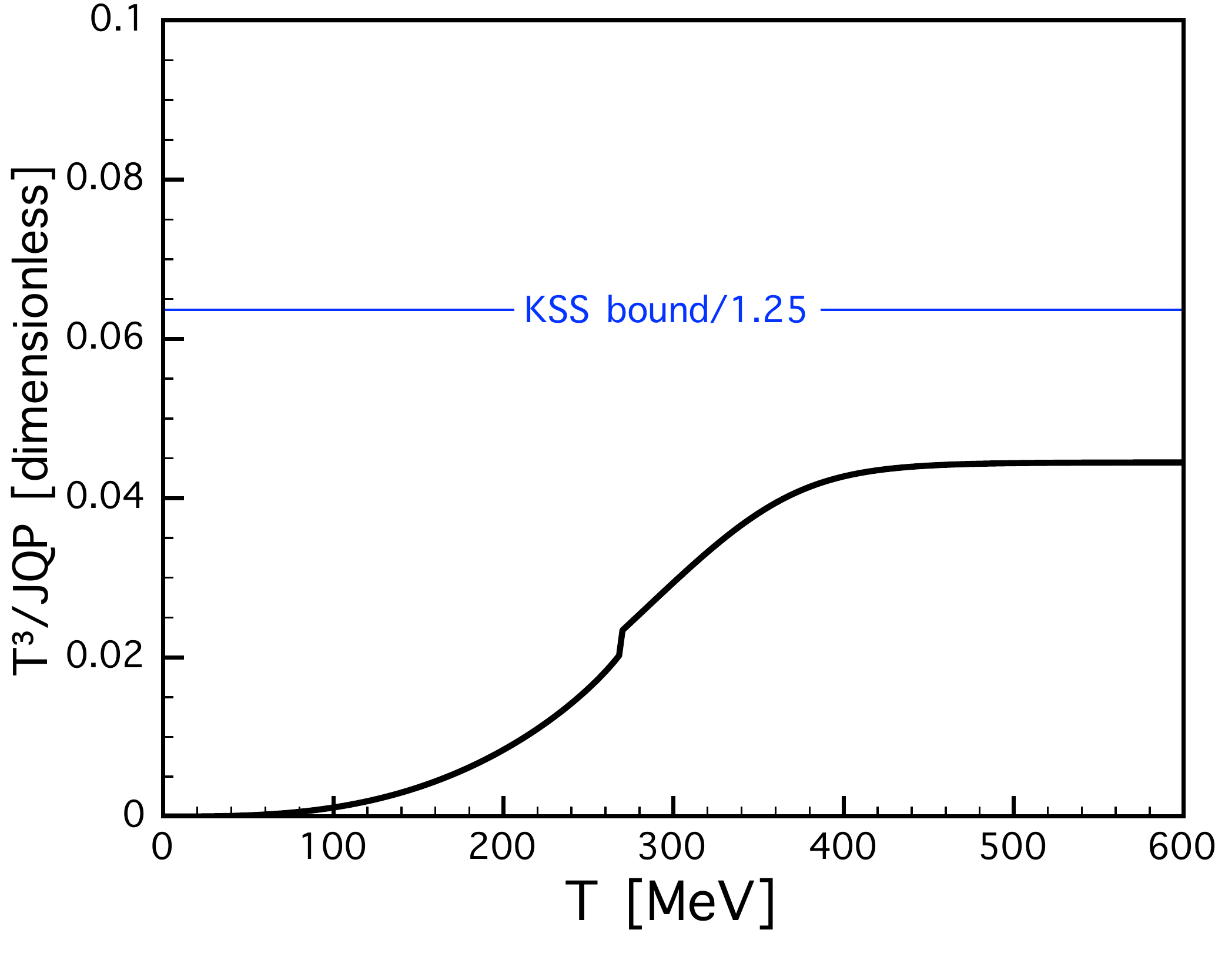}
\end{tabular}
\caption{(Color online) Right: Jet-quenching parameter (JQP) $\hat{q}$ in Eq.~(\ref{eq:JQP}) as a function of $T$. Left: $T^3/\hat{q}$ as a function of $T$. See the text for details.}       
\label{FIG34}
\end{figure}

%-------------------------------------------------
%\section{Numerical results and discussions}
%-------------------------------------------------
Now, we are in a position to present the numerical results for $\hat{q}$ with relevant discussions. In the left panel of Fig.~\ref{FIG34}, we present the numerical results separately for each $\hat{q}_n$ in Eq.~(\ref{eq:JQP}), in addition to the total result. We note that $\hat{q}_0$ with the gluon condensate (dot) is almost negligible in comparison to others. Similarly, $\hat{q}_1$ corresponding to the string tension (long-dash) gives only small contribution and is terminated at $T\approx280$ MeV by construction from the LQCD data, as depicted in the left panel of Fig.~\ref{FIG12}. In contrast, $\hat{q}_2$ (short-dash) and $\hat{q}_3$ (dot-dash), which are proportional to the Coulomb and constant potentials in Eq.~(\ref{eq:EFFPOT}), respectively, present dominant contributions in the opposite way, and their delicate competition makes difference, being led to the total $\hat{q}$ (thick-solid). We notice an increasing curve for $\hat{q}$ with respect to $T$ and come by numerically that $\hat{q}=(5\sim25)\,\mathrm{GeV/fm}$ for $T=(0\sim0.6)$ GeV. The parameterization of $\hat{q}$ for $T\to\infty$ in Eq.~(\ref{eq:PARAJQP}) is also drawn in the thin-solid line. We also demonstrate other theoretical results: The narrow diamond, inverted-triangle, square, diamond, circle, and triangle symbols designate the theoretical results from the NLO three-dimensional fluid dynamics~\cite{Majumder:2007ae}, stochastic QCD~\cite{Antonov:2007sh}, SU($2_c$) LQCD simulation~\cite{Majumder:2012sh},  AdS/CFT with the case-II hyperscaling violation~\cite{Sadeghi:2013dga}, AdS/CFT for $\mathcal{N}=4$ supersymmetric QCD~\cite{Liu:2006ug}, and AdS/CFT with chemical potential~\cite{Lin:2006au}. The shaded area stands for the time-averaged $\hat{q}$, extracted from the RHIC experiment~\cite{Eskola:2004cr,Dainese:2004te}. By comparing ours with other results, it turns out that our numerical results are well consistent with those from the AdS/CFT results, whereas the LQCD and other theoretical results are relatively smaller than ours. This observation tells us that our numerical result for $\hat{q}$ becomes proportional to $T^3$ as $T$ increases, as explicitly shown in Eq.~(\ref{eq:PARAJQP}), being similar to that  all the AdS/CFT estimations do as $\hat{q}_\mathrm{SYM}=\frac{\pi^2\sqrt{\lambda}}{a}T^3$~\cite{Liu:2006ug}. For $T=(300\sim500)$ MeV, our result agrees with the time-averaged $\hat{q}$ of RHIC~\cite{Eskola:2004cr,Dainese:2004te}. In Ref.~\cite{Majumder:2007zh}, it was discussed that the relation between the ratio $T^3/\hat{q}$ defines the strong- or weak-coupling region with the shear viscosity-to-entropy density ratio $\eta/s$ as follows:
%EQUATION>>>
\begin{equation}
\label{eq:STRONGWEAK}
\frac{\eta}{s}\,(\approx,\gg)\,1.25\frac{T^3}{\hat{q}}:\,\,\mathrm{(weak,strong)\,coupling}.
\end{equation}
%EQUAITON<<<
Note that there appeared the Kovtun-Son-Starinets (KSS) bound, signaling the QGP as an almost perfect fluid~\cite{Kovtun:2004de}: $\frac{\eta}{s}\ge\frac{1}{4\pi}$. We evaluate $T^3/\hat{q}$ and show the results in the right panel of Fig.~\ref{FIG34}. We also depict the KSS bound$/1.25$ in the figure (thin-solid). It turns out that the curve (thick-solid) increases up to $T\approx400$ MeV, then gets saturated to $\sim4.45\times10^{-2}$ beyond it. Again, this tendency indicates that $\hat{q}\propto T^3$ at the high-$T$ region.  In comparison to the KSS bound, our values are about $(30\sim50)\%$ smaller, so that the $T$ region, which we are interested in with the instanton framework, denotes possibly the strongly-coupled QGP as expected. 

%FIGURE>>>
%FIGURE<<<

%-------------------------------------------------
%\section{Summary and future perspectives}
%-------------------------------------------------
We make a brief summary: The jet-quenching parameter $\hat{q}$ has been scrutinized for $N_c=3$ in the nonperturbative region, employing the effective static (heavy) quark-antiquark potential with the help of the LQCD data and $T$-dependent instanton length parameters, providing a neat expression $\hat{q}\approx8V_{\bar{Q}Q}/L^3$ and showing the tendency $\hat{q}\propto T^3$ for high $T$. We arrived at an approximated expression for it, i.e. $\hat{q}\approx177.743\,T^3\,[\mathrm{GeV/fm}]$ as $T$ goes infinity. The numerical results are well consistent with other theoretical and experimental analyses, especially from those from AdS/CFT. It also turned out that the Coulomb and constant potentials of $V_{\bar{Q}Q}$ play a most important role to produce $\hat{q}$. Now, we are planing to derive the relation between $\eta/s$ and $T^3/\hat{q}$ in a consistent way to see whether the system is indeed characterized by the strongly-coupled QGP. Related works are under progress and will appear elsewhere. 
%-------------------------------------------------
%\section{Acknowledgment}
%-------------------------------------------------
%-------------------------------------------------

%(^o^)
\end{document}